\documentclass[10pt]{article}
\usepackage{amsmath,amssymb,amsthm,amsfonts,amstext,amsopn,amsxtra,dsfont}

\newcommand\C{{\ensuremath {\mathbb C} }}

\newcommand\id{{\ensuremath {\mathds 1} }}
\newcommand\Hh{{\mathcal{H}}}
\renewcommand\phi{\varphi}
\newcommand\Tr{{\rm Tr\,}}
\newcommand\half{\mbox{$\frac 12$}}

\begin{document}
\title{On the failure of subadditivity of the Wigner-Yanase entropy}
\author{\vspace{5pt} Robert Seiringer
\\ \vspace{-4pt}\small{ Department of Physics, Jadwin Hall, Princeton
University}\\ \vspace{-4pt}\small P.O. Box 708, Princeton, NJ
08542, USA
\\ {\small Email: \texttt  {rseiring@math.princeton.edu}} }

\date{April 3, 2007}
\maketitle

\begin{abstract}
  It was recently shown by Hansen that the Wigner-Yanase entropy is,
  for general states of quantum systems, not subadditive with respect
  to decomposition into two subsystems, although this property is
  known to hold
  for pure states. We investigate the question whether the weaker
  property of subadditivity for pure states with respect to
  decomposition into more than two subsystems holds. This property
  would have interesting applications in quantum chemistry. We show,
  however, that it does not hold in general, and provide a
  counterexample.
\end{abstract}

\renewcommand{\thefootnote}{${\,}$}
\footnotetext{Work partially supported by U.S. National Science
Foundation grant PHY-0353181 and by an Alfred P. Sloan Fellowship.}
\renewcommand{\thefootnote}{${\, }$}
\footnotetext{\copyright\,2007 by the author.
This paper may be reproduced, in its entirety, for non-commercial
purposes.}

\bigskip

In 1963, Wigner and Yanase \cite{WY1} introduced the entropy-like quantity
\begin{equation}
S^{\rm WY}(\rho,K)= \half\Tr[\rho^{1/2},K]^2= \Tr\rho^{1/2}K\rho^{1/2}K -\Tr \rho K^2 
\end{equation}
for density matrices $\rho$ of quantum systems, with $K$ some fixed
self-adjoint operator. They showed that $S^{\rm WY}$ is concave in
$\rho$  \cite{WY1,WY2} and, for pure states, subadditive with respect to decomposition
of the quantum system into two subsystems. More precisely, if
$|\psi\rangle$ is a normalized vector in the tensor product of two
Hilbert spaces, $\Hh_1\otimes \Hh_2$, and $K_1$ and $K_2$ are
self-adjoint operators on $\Hh_1$ and $\Hh_2$, respectively, then
\begin{equation}
  S^{\rm WY}(|\psi\rangle\langle\psi|,K_1\otimes\id 
+ \id\otimes K_2) \leq S^{\rm WY}(\rho_1,K_1) + S^{\rm WY}(\rho_2,K_2)\,,
\end{equation}
where $\rho_1=\Tr_{\Hh_2}|\psi\rangle\langle\psi|$ and
$\rho_2=\Tr_{\Hh_1}|\psi\rangle\langle\psi|$ denote the reduced states
of the subsystems.  Recently, it was shown by Hansen \cite{hansen}
that this subadditivity {\it fails} for general mixed states.

This leaves open the question whether the Wigner-Yanase entropy is
subadditive for pure states with respect to decompositions into {\it more}
than 2 subsystems. If true, this property would have interesting
consequences concerning density matrix functionals used in quantum
chemistry, as will be explained below. We shall show, however, that
this property does {\it not} hold, in general.

Let $\rho=|\psi\rangle\langle\psi|$ be a pure state on a tensor
product of $N$ Hilbert spaces, $\Hh=\bigotimes_{i=1}^N \Hh_i$, and let
$K_i$ be self-adjoint operators on $\Hh_i$. For simplicity we use the
same symbol for the operators on $\Hh$ which act as the identity on
the remaining factors. Subadditivity of $S^{\rm WY}$ would mean that
\begin{align}\nonumber
 - S^{\rm WY}(|\psi\rangle\langle\psi|,\mbox{$\sum_i K_i$}) & = \left\langle\psi\left|
      \left(\mbox{$\sum_i K_i$}\right)^2 \right|\psi\right\rangle -
  \left\langle\psi\left| \mbox{$\sum_i K_i$}\right|\psi\right\rangle^2
  \\ & \geq \sum_i \left( \Tr_{\Hh_i} \rho_i K_i^2 - \Tr_{\Hh_i} \rho_i^{1/2}K_i
    \rho_i^{1/2}K_i\right) 
\,, \label{sn}
\end{align}
where $\rho_i$ is the reduced density matrix of $|\psi\rangle\langle\psi|$
on $\Hh_i$.

Assume now that all the $\Hh_i$ are equal to the same $\Hh_1$, say, and
that also all the $K_i$ are equal, i.e., $K_i$ acts as $K$ on the
$i$'th factor for some fixed operator $K$ on $\Hh_1$. Ineq.~(\ref{sn}) together with
concavity of $S^{\rm WY}$ would thus imply that
\begin{equation}\label{val}
  \left\langle\psi\left|
      \left(\mbox{$\sum_i K_i$}\right)^2 \right|\psi\right\rangle -
  \left\langle\psi\left| \mbox{$\sum_i K_i$}\right|\psi\right\rangle^2 
\geq \Tr_{\Hh_1} \gamma K^2 -\Tr_{\Hh_1} \gamma^{1/2}K\gamma^{1/2}K\,,
\end{equation}
or
\begin{equation}\label{vali}
\left\langle\psi\left|
      \mbox{$\sum_{i\neq j} K_i K_j$} \right|\psi\right\rangle 
\geq \left(\Tr_{\Hh_1}\gamma K\right)^2  -\Tr_{\Hh_1} \gamma^{1/2}K\gamma^{1/2}K\,,
\end{equation}
where $\gamma=\sum_i \rho_i$ denotes the one-particle density matrix
of $|\psi\rangle\langle\psi|$. This represents a {\it correlation
  inequality}, bounding from below two-particle terms in terms of
one-particle terms only.

As explained in \cite{FLSS}, the validity of (\ref{val}) for
continuous quantum systems in the case where $K$ is the characteristic
function of a ball of arbitrary size and location would imply that the
ground state energies of Coulomb systems like atoms and molecules could
be bounded from below by a density-matrix functional introduced by
M\"uller \cite{muller}. For $N=2$ this follows from the result in \cite{WY1}.

\renewcommand{\thefootnote}{$1$}

In the following, we shall show that, in general, (\ref{val}) fails to
hold for $N=3$, and hence for all $N\geq 3$. We choose the simplest nontrivial three-particle 
Hilbert space, $\C^2\otimes \C^2\otimes \C^2$, and pick a basis
$\{|\uparrow\rangle,|\downarrow\rangle\}$ in $\C^2$. We choose
$K=|\uparrow\rangle\langle\uparrow|$, and\footnote{This particular counterexample was found
  with the aid of the computer algebra software {\it Mathematica}.}
\begin{align}\nonumber
\psi(\uparrow,\uparrow,\uparrow) &= \frac 2{\sqrt{55}} \\ \nonumber
\psi(\uparrow,\uparrow,\downarrow) =\psi(\uparrow,\downarrow,\uparrow)=\psi(\downarrow,\uparrow,\uparrow) & =\frac 4{\sqrt{55}} \\ \nonumber
\psi(\uparrow,\downarrow,\downarrow) =\psi(\downarrow,\uparrow,\downarrow)=\psi(\downarrow,\downarrow,\uparrow) & =\frac 1{\sqrt{55}} \\
\psi(\downarrow,\downarrow,\downarrow) &=0\,.
\end{align}
\
Then
\begin{align}\nonumber
\langle\psi|\psi\rangle&=\frac 1{55}\left( 2^2+3*4^2+3*1\right)=1 \\ \nonumber
\left\langle\psi\left| \mbox{$\sum_i K_i$}\right|\psi\right\rangle &= \frac{1}{55}\left( 3*2^2+2*3*4^2+1*3*1\right)= \frac{111}{55}\\
\left\langle\psi\left| \left( \mbox{$\sum_i K_i$}\right)^2\right|\psi\right\rangle &= \frac{1}{55}\left( 3^2*2^2+2^2*3*4^2+1*3*1\right)= \frac{231}{55}
\end{align}
and hence the left side of Ineq.~(\ref{val}) equals
\begin{equation}
\frac{231}{55}-\left(\frac{111}{55}\right)^2 =\frac{384}{3025}\approx 0.126942\,. 
\end{equation}

The one-particle density matrix $\gamma$ is given by the $2\times 2$-matrix
\begin{equation}
\gamma = \frac 3{55} \left( \begin{array}{cc} 37 & 16 \\ 16 & 18 \end{array}\right)
\end{equation}
whose square root equals
\begin{equation}
\gamma^{1/2} \approx \sqrt{\frac 3{55}} \left( \begin{array}{cc}
    5.85827 & 1.63729\\ 1.63729 & 3.91399 \end{array}\right)\,.
\end{equation}
Hence the right side of (\ref{val}) is
\begin{equation}
\frac{3}{55} \left( 37 - (5.85827)^2\right) \approx  0.146221 > 0.126942\,. 
\end{equation}
This shows that Ineq. (\ref{val}) fails in general for $N>2$, and
hence the Wigner-Yanase entropy is not subadditive with respect to the
decomposition of pure states into more than $2$ subsystems.

We note that the same counterexample can also be constructed for
continuous quantum systems, where $K$ equals the characteristic functions of
some measurable set $B$. One simply takes $B$ and $\Omega$ to be two
disjoint sets, each with volume one, and sets
\begin{equation}\label{psic}
\psi(x_1,x_2,x_3) = \frac 1{\sqrt{55}}\left\{ \begin{array}{ll}
2 & {\rm if\ all\ 3\ particles\ are\ in\ }B \\
4 & {\rm if\ 2\ particles\ are\ in\ } B{\rm \ and\ 1\ in\ \Omega}\\
1 & {\rm if\ 1\ particle\ is\ in\ }B{\rm \ and\ 2\ in\ \Omega}\\
0 & {\rm otherwise.}
\end{array}\right.
\end{equation}
This leads to the same counterexample as above.

Similarly, one can construct a counterexample for fermionic (i.e.,
antisymmetric) wavefunctions which, after all, is the case of interest
in \cite{FLSS}. Simply take $(x,y)$ as the coordinates of one
particle, choose the wave function to be the product of (\ref{psic})
for the $x$ variables and a Slater-determinant for the $y$ variables,
which is non-zero only if all the $y$'s are in some set $\Lambda$. If
$K$ denotes multiplication by the characteristic function of
$B\times \Lambda$, this leads to the same counterexample as
before.

\end{document}